\documentclass[
aps,nofootinbib,
showpacs,showkeys,preprint
tightenlines,preprintnumbers,] {revtex4}

\usepackage{epsf,epsfig,subfigure,graphicx,amsmath,amssymb 
}
\usepackage{color}
\newcommand{\dis}[1]{\begin{equation}\begin{split}#1\end{split}\end{equation}}
\newcommand{\ie}{{\it i.e.~}}

\newcommand{\UPQ}{U(1)$_{\rm PQ}$}
 
\def\sw0{{$\sin^2\theta_W^0$}}

\newcommand{\Z}{{\bf Z}}

\def\E6{{\rm E_6}}

\def\EE8{{\rm E_8\times E_8'}}

\begin{document}

\draft

\title{Hairs of discrete symmetries and gravity} 
 
\author{Kang Sin Choi}
\address
{Scranton Honors Program, Ewha Womans University, Seodaemun-Gu, Seoul 03760, Republic of Korea}

\author{Jihn E.  Kim}
\address
{Department of Physics, Kyung Hee University, 26 Gyungheedaero, Dongdaemun-Gu, Seoul 02447, Republic of Korea, and\\
Center for Axion and Precision Physics Research (IBS),
  291 Daehakro, Yuseong-Gu, Daejeon 34141, Republic of Korea
}

\author{Bumseok Kyae}
\address
{Department of Physics, Pusan National University, 2 Busandaehakro-63-Gil, Geumjeong-Gu, Busan 46241, Republic of Korea}

\author{Soonkeon Nam}
\address
{Department of Physics, Kyung Hee University, 26 Gyungheedaero, Dongdaemun-Gu, Seoul 02447, Republic of Korea}

\begin{abstract} 
Gauge symmetries are known to be respected by gravity because  gauge charges carry flux lines, but global charges do not  carry flux lines and are not conserved by gravitational interaction. For discrete symmetries, they are spontaneously broken in the Universe, forming domain walls. Since the realization of discrete symmetries in the Universe must involve  the vacuum expectation values of Higgs fields, a string-like configuration (hair) at the intersection of domain walls in the Higgs vacua can be realized. Therefore, we argue that discrete charges are also respected by gravity.
\end{abstract}
\maketitle


\section{Introduction}\label{sec:Introduction}
 It has been known for a long time that  discrete sub-groups of gauge groups, the so-called discrete gauge symmetries, are not broken by gravitational interactions   \cite{KraussWil88,Banks89}. Effects of quantum gravity  are studied by looking at various topologies of the metric tensor $g_{\mu\nu}$. If some gauge charges are separated from our Universe by metric change, the separated gauge charges cannot be completely hidden from our Universe because they leave long range flux lines. On the other hand, if global charges are separated from our Universe, the lost charges leave no hint to an observer in our Universe and he notices that global charges are not conserved in our Universe. Thus,  gauge symmetries are not broken but global symmetries are broken by metric changes.
 This is the basic reasoning that discrete gauge symmetries are used in particle physics \cite{Ibanez92}.  This top-down approach on discrete symmetries fits to the string compactification \cite{Koba07,KimPLB13} because string theory does not allow any global symmetry. 
 
In the bottom-up approach,  the flux line argument is not so clear. It uses just the classical gauge fields and does not rely on the renormalizability in the theory of elementary particles. To be specific, let us consider a continuous symmetry U(1). If U(1) is a gauge symmetry, it does not have any gauge anomaly. If U(1) is a global symmetry, it may have a gauge anomaly U(1)--$G-G$ where $G$ is a gauge group as in the Peccei-Quinn (PQ)  global symmetry \UPQ\,\cite{PQ77}. Obstructing the PQ symmetry needed for an ``invisible'' axion was based on this argument \cite{Barr92}.

However, the absence of any gauge anomaly is not a guarantee for a gauged U(1) symmetry.  Some global U(1) symmetries may not have any gauge anomaly. The difference in the gauge and global symmetries resides in the property on the local transformation, \ie using  a covariant derivative  $D_\mu=\partial_\mu-i A_\mu$ in gauge theories, or just an ordinary one $\partial_\mu$ in global symmetries. A discrete subgroup of U(1) cannot know whether the mother U(1) is gauged or not.  In the bottom-up approach, there must be some other reason for the effects of the metric change.
 
 In this paper, we adopt the concept of ``hair'' which   means that hair's thickness is the same at any distance from the surface of the head. At the surface, there must be fields at the surface for a hair to be defined. This definition excludes any possibility for hairs of global symmetries. In gauge theories, there are gauge fields at the surface.
In gauge theories, the relation of the fields at the surface with the charge $Q$  in the volume enclosed by the surface is provided by the equations of motion and current conservation.  Existence of hairs is crucial in guaranteeing the symmetry in the presence of the gravitational interaction. It is known that black holes have gauge-charge hairs, 
which will be briefly commented in parallel with our method.
  
For a gauge charge $Q$, we have gauge fields spreading out from $Q$. Consider the current $j^\mu$ and the corresponding electric field {\bf E} along a line to be interpreted as a hair. We can perform local transformations such that  {\bf E} is the same along a line but zero outside the line, which behaves as a hair.\footnote{Here, the line is not a mathematical one but has some physical thickness. 
Thus, gauge charges can have hairs but global charges cannot, and metric changes know only hairs.} 

In this paper, we show how discrete charges can have hairs in the bottom-up approach, and derive that discrete symmetries are not broken by gravity.    For an explicit presentation, we will present examples with the Abelian discrete symmetry $\Z_N$ and in particular with $\Z_2$ illustrations.

\section{Discrete charges of $\Z_N$ vacua}\label{sec:DWvac}
\begin{figure}[!t]
\begin{center}
\includegraphics[width=0.6\linewidth]{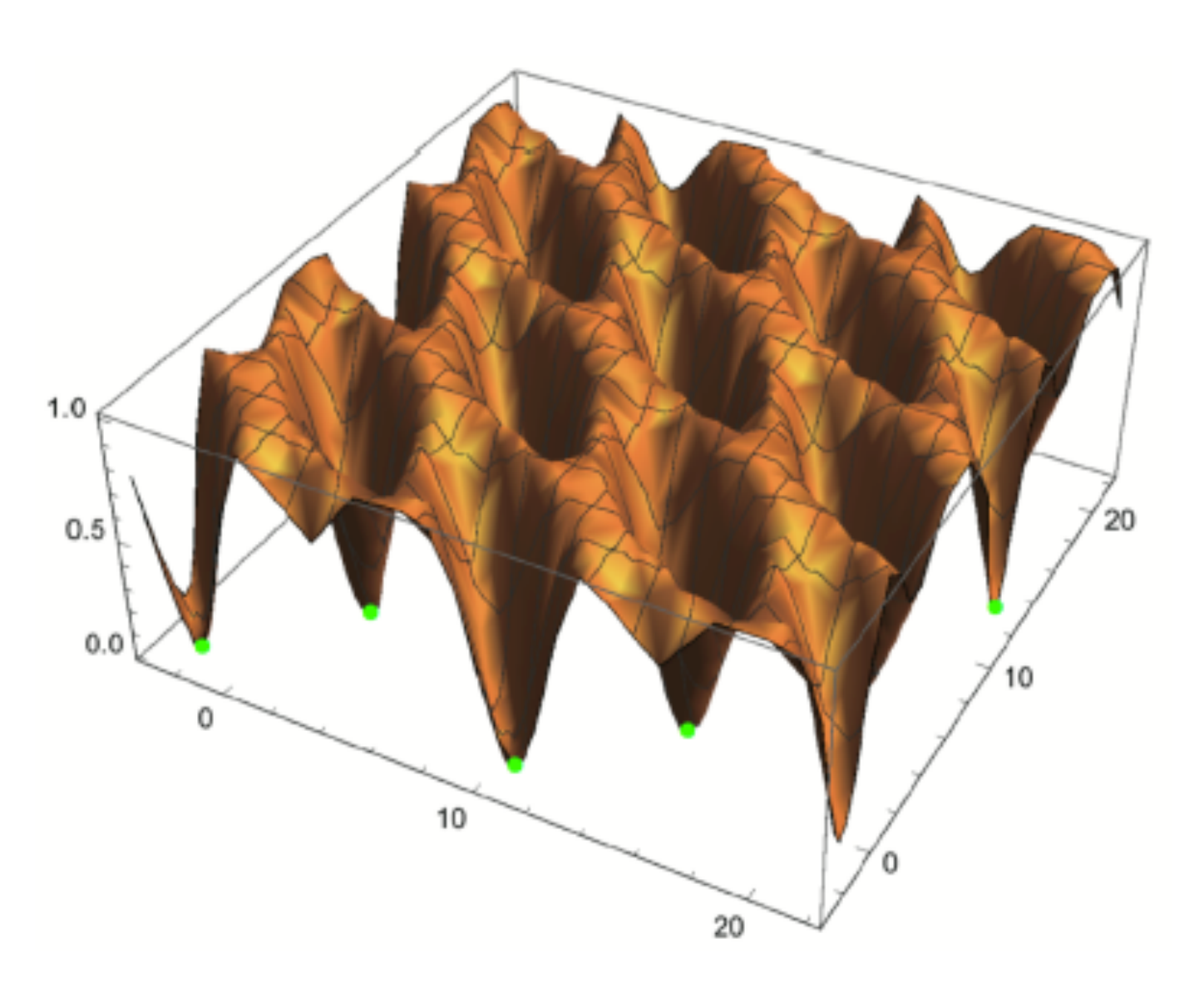}
\end{center}
\caption{Multiple discrete vacua. Some of minima are shown as green bullets. } \label{fig:Vacua}
\end{figure}

A discrete symmetry is defined by the number of minima of the potential $V$ such as in Fig. \ref{fig:Vacua}. Let us consider one minimum, say a green bullet in Fig. \ref{fig:Vacua}. We can choose the value of the Higgs field to be zero at that point so that the discrete symmetry is realized by the Wigner-Weyl manner. If it has a flat direction there, then one must consider a continuous symmetry, which has been spontaneously broken already. Not considering continuous symmetries, with the multiple vacua of  Fig. \ref{fig:Vacua}, the discrete symmetry is good at any point of the minima. We will consider the discrete charges at such a minimum.

\begin{figure}[!t]
\begin{center}
\includegraphics[width=0.6\linewidth]{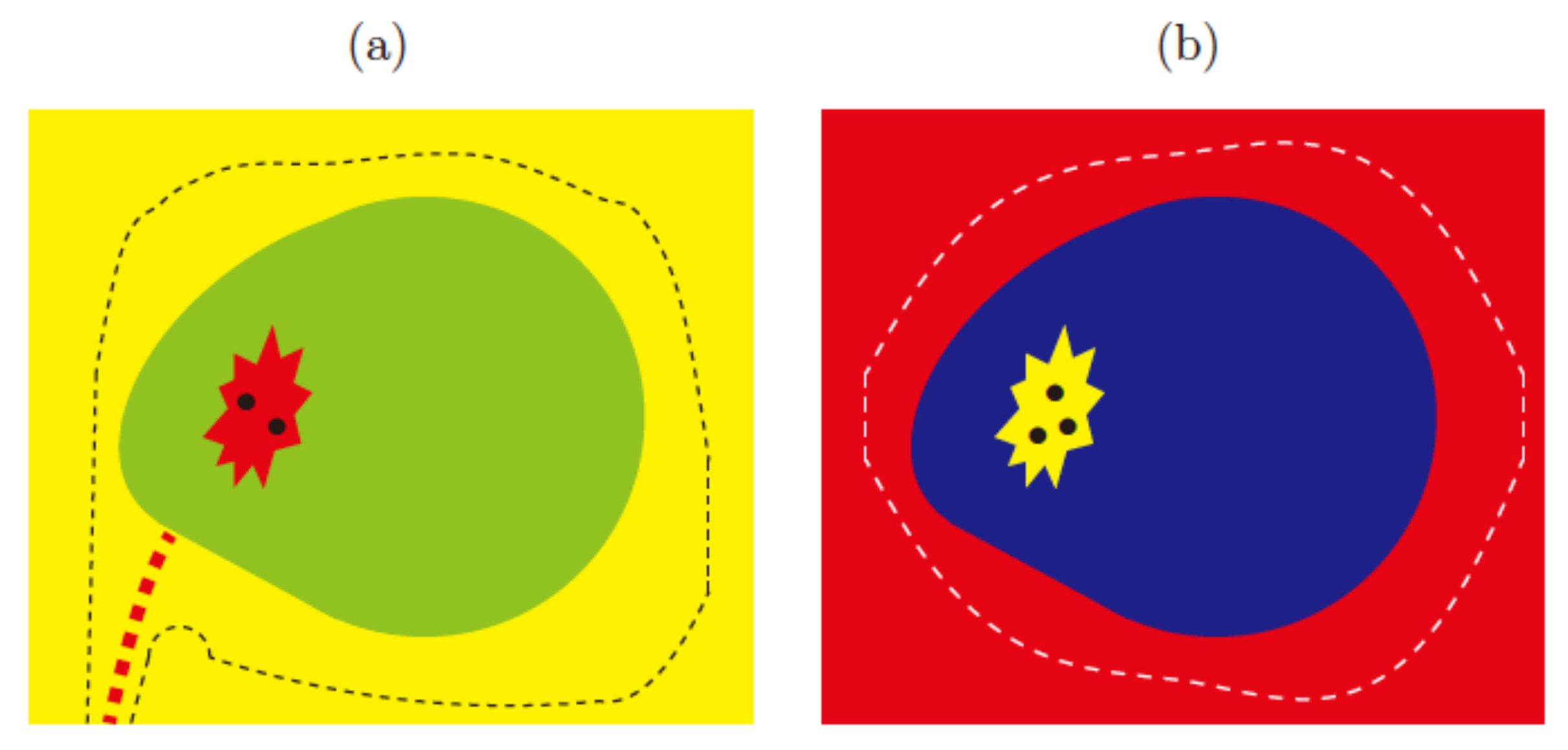} 
\end{center}
\caption{(a) A walled vacum (red, $q_{\rm total}=1$) seen from the $q=0$ (yellow) vaccum. Inside the wall, the opposite $q=1$ (red) vacuum is seen through a crack in the wall. This view of the wall is colored limegreen. (b) A walled vacum (yellow, $q_{\rm total}=0$) seen from the $q=1$ (red) vaccum.  This view of the wall is colored blue.  Dashes represent closing surfaces and black dots  represent particles. } \label{fig:DiscrVac}
\end{figure}

Realization of discrete symmetries in the Universe leads to domain walls \cite{Okun74}.
In the ``invisible'' axion case \cite{KimPRP87}, the Peccei-Quinn symmetry leads to $\Z_N$ domain walls \cite{Sikivie82}. For the Kim-Shifman-Vainstein-Zakharov ``invisible'' axion where there is only one vacuum \cite{KSVZ}, even the $\Z_1$ domain wall can be considered in the Universe evolution \cite{VilenkinE82}. In this case, however, all space points except at the wall are in the same vacuum. Different vacua arise for the cases of $N\ge 2$. Two kinds of walled vacua are possible for  $\Z_2$, viz. Fig. \ref{fig:DiscrVac}. Two vacua of $\Z_2$ are defined with discrete charges   $q=2n$ and $2n+1,$ mod. 2 ($n$\,=\,integer). 

In Fig. \ref{fig:DiscrVac}\,(a), the (red, $Q_{\rm total}=1$) vacum  is seen from the $q=0$ (yellow) vaccum. A closed domain wall separates these two. This wall viewed from the yellow vacuum is symbolized by the limegreen color.  In Fig. \ref{fig:DiscrVac}\,(b), the $q=0$ (yellow, $Q_{\rm total}=0$) vacum  is seen from the $q=1$ (red) vaccum. The wall viewed from the red vacuum is symbolized by the blue color. In Fig. \ref{fig:DiscrVac}\,(a),  the dashed boundary encloses the walled  $q=1$ vacuum. A scalar field $\phi$ in the $q=0$  or $q=1$ vacua is represented by 
$e^{iq\pi}R({\bf x})$. 

Let us illustrate examples in $\Z_2$. Then, $q$ can be 0 or 1. 
For $q=0$, we use the field VEV $\phi=0$.\footnote{If $\phi=- v $ corresponds to $q=0$,  we add a constant $v$ to simplify the value of $\phi$.} For a  ball of discrete charge $q$, the radius of the ball is determined by minimizing the energy
   \begin{equation}
E_\omega= E+\omega\left[q-\frac{1}{2i}\int d^3x \,(\Phi^*\partial_t\Phi - \Phi\partial_t\Phi^* )\right],\label{eq:Qchar}
\end{equation}
where $\omega$ (with the energy dimension) is the Lagrange multiplier and $q$ of the ball can be defined as
    \begin{equation}
q= \frac{1}{2i}\int d^3x \,,(\Phi^*\partial_t\Phi - \Phi\partial_t\Phi^* ).\label{eq:Qcharge}
\end{equation}
   
In the evolving Universe, the vacuum inside the ball expands such that $\phi$ in the red becomes constant.
For a spherically symmetric $R({\bf x})$, let us parametrize it as
 \begin{equation}
\Phi=  \sqrt{\frac{3k^3}{4\pi^4 \omega}}\,e^{i\omega t}\left\{
\begin{array}{l}
1,~~\textrm{for~~}0\le r<\frac{\pi}{k}\\[0.5em] 0,~~\textrm{for~~}r>\frac{\pi}{k}
 \end{array} \right.\label{eq:FieldValue}  
 \end{equation}
 where $\pi/k$ is the radius of the ball,
and  $\Phi=0$   in the yellow part of Fig. \ref{fig:DiscrVac}\,(a). So, we obtain
\begin{equation}
\frac{1}{2i}\int_{\rm (inside\,dashed)} d^3x\,(\Phi^*\partial_t\Phi - \Phi\partial_t\Phi^* ) =1.\nonumber
\end{equation}
The total charge $q$ inside the dashed surface of Fig. \ref{fig:DiscrVac}\,(a) is 1, and the dashed string symbolyzes this fact. 

Definition of charge $q$ by Eq. (\ref{eq:Qcharge}) is not by the N$\ddot{\rm o}$ther current. It is simply defined by the vacuum expectation value (VEV) of the phase of a Higgs field $\Phi$. To relate this charge $q$ to the charge defined by the   N$\ddot{\rm o}$ther current, the $t$ dependence of $\Phi$ is introduced as the example in Eq. (\ref{eq:FieldValue}). To make it an integer, the VEV which is designed as a constant\footnote{In   the connected portion in the Universe, the minimum of the potential with a fixed value of $\Phi$ is chosen everywhere.} is appropriately chosen. Equation (\ref{eq:Qchar}) is the matching condition to the charge $q$ calculated by the  N$\ddot{\rm o}$ther current. Discrete symmetries in the Universe are realized by the VEVs of Higgs field $\Phi$ having degenerate minima as shown in Fig. \ref{fig:Vacua}. So, it is appropriate to figure out the discrete charges in the vacuum of spin-0 bosons as shown in Fig. \ref{fig:DiscrVac} where the limegreen surface separates two different Higgs portions in the Universe.      In case of  Fig.  \ref{fig:DiscrVac}\,(b), we apply discrete transformation $e^{i\pi Q}$. Then it is equivalent to the discussion in  Fig.  \ref{fig:DiscrVac}\,(a). 

In  Fig. \ref{fig:DiscrVac}\,(b), $q_{\rm total}=4={\rm even}=0$ within the white dashed surface, which must be the case for any closed surface. So, the white dashed surface of Fig. \ref{fig:DiscrVac}\,(b) can be shrunk to naught. In  Fig. \ref{fig:DiscrVac}\,(a), $q_{\rm total}=3={\rm odd}=1$ within the  dashed surface, which must be the case for any closed surface, which encloses the limegreen ball.  To identify $q_{\rm total}$, we extend a dash line from the limegreen ball up to the horizon.
 The dashed surface of Fig.  \ref{fig:DiscrVac}\,(a) is  named as an infinite-tail ``tadpole'', having a head and a long tail.
  At the limegreen surface, there exists the Higgs field which provides the logistics for a hair to exist in our case.
 
Suppose that the U(1)  charges of $\Phi$ and $\psi$ are 2 and 1, respectively, and there is no U(1)$-G-G$ anomaly where $G$ is a gauge group. Then, this U(1) can be a gauge or global group. The U(1) transformations by an angle $\theta$ on the fields are  $\Phi\to e^{i\,4\pi  \theta}\Phi$ and $\psi\to e^{i\,2\pi  \theta}\psi$.
A potential invariant under U(1) is
\begin{eqnarray}
V=-\left(\sqrt{2}\mu\,\psi^{*\,2}\Phi+\textrm{h.c.}\right)+ \cdots
\label{eq:V}
\end{eqnarray}
where $\cdots$ are other U(1)  invariant terms. At a high scale $f$, the U(1) is broken to $\Z_2$ by the VEV, $\langle \Phi\rangle=f/\sqrt2$. Let us split $\psi$ to a radial and phase fields,
\begin{eqnarray}
\psi=\frac{v+\rho}{\sqrt2}\,e^{i\,2\pi\theta}
\label{eq:psi}
\end{eqnarray}
  For a moment, let   $\langle v\rangle =0$ where $\theta=\phi/v$.
 At this stage, there is no $\theta $ dependence in the potential because $\langle \phi\rangle=0$. There is no way to distinguish the U(1) as a global or gauge group. Now, let $|\psi|$ develop  a VEV, $\langle |\psi|\rangle=v/\sqrt2$, so that the $\Z_2$ symmetric potential, $-(f\mu\,\psi^{*\,2} +\textrm{h.c.} )$, develops a $\theta$ dependence,
\begin{eqnarray}
V=-f\mu\,v^2\,\cos\theta+ \cdots\label{eq:WallEn}
\end{eqnarray}
In the region $\phi=-v$ (yellow) and $+v$ (red), let $q=0$ (yellow) and $1$ (red), respectively.\footnote{The discrete charges are the phase values of scalar fields.}
Now, the difference appears. In a gauged U(1), $\theta$ becomes redundant in the sense that it can be removed from $V$ by a gauge transformation \cite{Banks89}. But,  a global U(1) manifests itself in the form of the domain-wall energy density shown in Eq. (\ref{eq:WallEn}). 
The effective thickness of the domain wall is
\begin{equation}
\lambda_{\rm thickness}\approx 
\frac{\sqrt{\lambda}}{v}\,\sqrt{ \frac{f}{\mu}}\label{eq:thickness}
\end{equation}
where $\lambda$ is a quartic coupling constant, determining $f$ from the neglected terms in (\ref{eq:V}).
So, the domain wall can be seen as we discussed above, and the intersection of domain walls works as a hair.

\section{Discrete charge conservation in the bottom-up approach}

A conserved current of a continuous symmetry is constructed with an infinitesimal shift of fields, $\Psi\to e^{i\,2\pi\varepsilon(x) Q}\Psi$ where $Q$ is the generator of the transformation. For  a discrete charge $I$ of $\Psi$ in $\Z_N$ symmetry, we adopt this around a specific discrete value $I$ with $I=\{1,2,\cdots,N-1\}$,  
\begin{eqnarray}
 \Psi\to   e^{i\,2\pi\left[\frac{I}{N}+\varepsilon(x) \right]}\Psi. \label{eq:ChargeI}
\end{eqnarray}
The vacua of fields are defined for specific values of scalar fields as in Sec. \ref{sec:DWvac}, which is related to the charges of quanta, Eq. (\ref{eq:ChargeI}), in the volume via  conservation of the current.   
 
\begin{figure}[!t]
\begin{center}
 \includegraphics[width=0.7\linewidth]{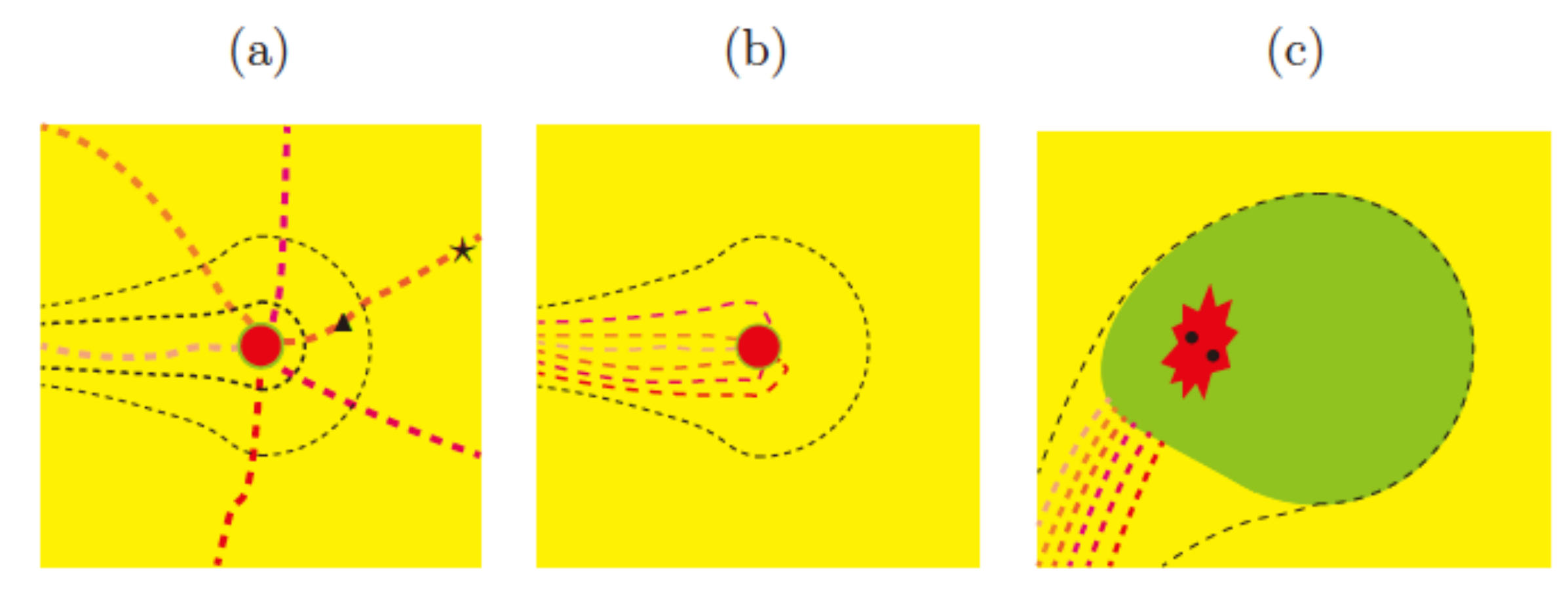}
\end{center}
\caption{(a)  A $\Z_N$ walled ball seen in the $q=0$ vacuum. The tail of the ``tadpole'' can have $n=0,1,\cdots,N-1$ dashed lines. $N$ dashed line is equivalent to no tail, \ie no $\Z_N$ charge. (b)  Another view of (a) by the discrete flux, and (c) an expanded view of (b) with the dashed boundary touching the wall.  In (a),  the discrete charges inside two dashed surfaces are the same. } \label{fig:DiscFlux}
\end{figure}

For a discrete symmetry, the vacuum structure of Higgs fields allows a hair(s) as shown in Sec. \ref{sec:DWvac}. This string-like hair starts from a nonzero discrete charge $q$ and  ends at another discrete charge or at the horizon. In Fig. \ref{fig:DiscFlux}\,(a), we show $\Z_N$ hairs spreading out from a nonzero discrete charge. The Higgs vacua are defined by the VEVs of Higgs fields. In Fig. \ref{fig:DiscFlux}\,(a), we enclose the discrete charges by two dashed surfaces, the thick and thin ones. The discrete charges inside the thin and thick surfaces are exactly the same. If we assign one discrete charge to one dashed line, the discrete charges inside the thin and thick surfaces are exactly the same, \ie the hair interpretation of the dashed lines equates these two estimations. For example, the Higgs vacuum value at the star at the dashed line is the same as that at the triangle.  Calculating the discrete charge in \ref{fig:DiscFlux}\,(b) is like calculating the discrete charge in  Fig. \ref{fig:DiscFlux}\,(c) where we made it clear by moving the dashed surface touching the wall at the RHS of \ref{fig:DiscFlux}\,(c). So, when discrete charges move, we can consider them dragging dash lines corresponding to some units of discrete charges. 
 
\begin{figure}[!t]
\begin{center}
\includegraphics[width=0.6\linewidth]{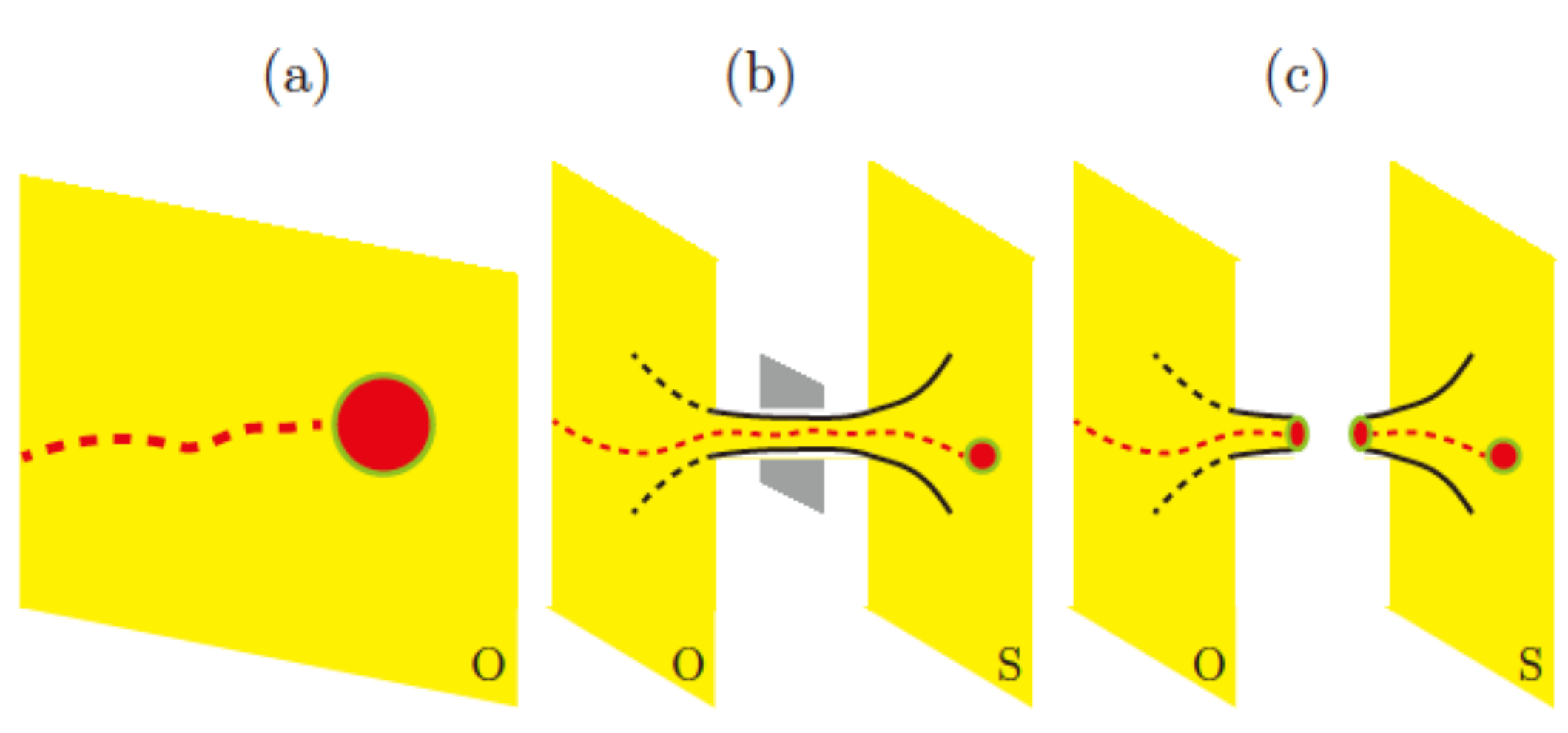} 
\end{center}
\caption{(a) A walled ball seen in the $q=0$ vacuum. The tail of a tadpole-like configuration extends to the horizon.  (b)  A tadpole passing through a wormhole. (c) The wormhole cut in the gray plane in (b). The observer O recovers the discrete charge. } \label{fig:Wormholes}
\end{figure}

Now let us proceed to discuss the wormhole effects in the metric theory of gravity.  Through wormholes, discrete charges can flow out from our Universe O. For simplicity, now let us focus on $\Z_2$. An infinite-tail  ``tadpole'' is symbolized in  Fig.  \ref{fig:Wormholes}\,(a), where  $Q_{\rm total}=1$.  Discrete charge  flow can be visualized as a ``tadpole'' passing through the wormhole as shown in Fig. \ref{fig:Wormholes}\,(b). If one tries to separate out the shadow world S from O by cutting the wormhole through the gray plane  in Fig. \ref{fig:Wormholes}\,(b), it cuts the infinite tail of ``tadpole''.   Then, at the cut plane, dashed lines are attached to a walled ball at each surface as shown in Fig.  \ref{fig:DiscrVac}\,(c). Recovering the wormhole throat, the observer O confirms that no discrete charge is lost.  To the observer O, gravitational effects do not break the discrete symmetry in consideration. This type of wormhole argument was used for a U(1) gauge symmetry in Ref. \cite{KimPLB13}.

\section{Scalar vacua}
 \begin{figure}[!t]
\begin{center}
\includegraphics[width=0.5\linewidth]{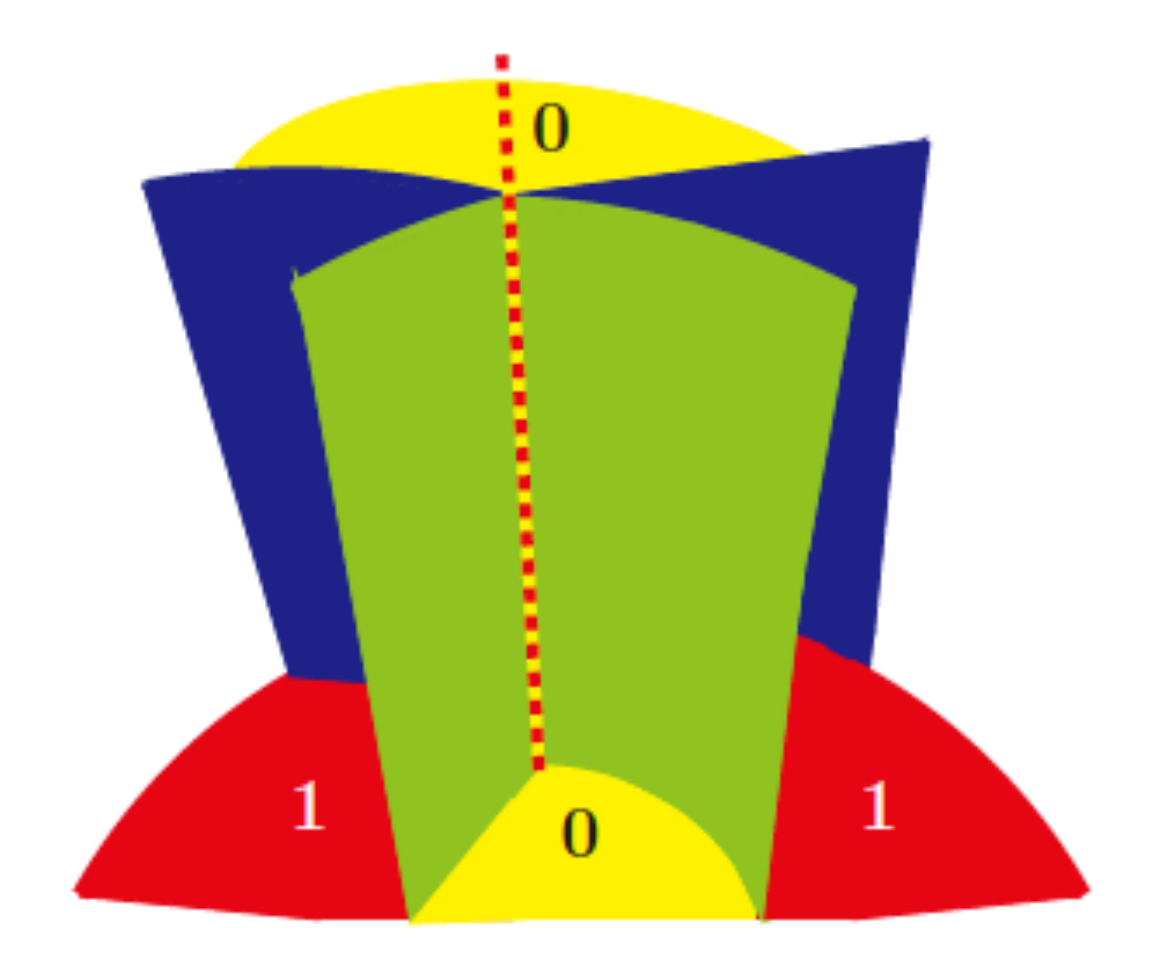} 
\end{center}
\caption{ A $\Z_2$ string. } \label{fig:TwoDdelta}
\end{figure}

Let us illustrate how such  hairs from discrete charges are set up in the spontaneously broken $\Z_N$ vacua. For simplicity, we present it in $\Z_2$. Let us consider a dashed surface of Fig. \ref{fig:DiscrVac}\,(a) . Our objective is to obtain a two dimensional delta function  at the surface of radius $r$ in the spherical-polar coordinate system, $(\theta_0,\varphi_0)$.  Consider  an effective $\Z_2$ symmetric action,
\begin{eqnarray}
{\cal L}_{\rm eff}=\frac{1}{M_{\rm eff}^2}\,\left(\partial ^\mu \partial^\nu\psi^*\right)\,\left({\partial}_\mu \partial_\nu\psi\right) .\label{eq:EffZ2}
\end{eqnarray}
At a closed surface, we can consider an effective Lagrangian where  the Lorentz symmetry is broken as in Fig. \ref{fig:DiscrVac}\,(a), and   contract  with the $(\mu i)$ indices in (\ref{eq:EffZ2}), $j^{\mu i}$. To have currents related to (\ref{eq:EffZ2}), consider a shift in the discrete vacuum $I$, viz. Eq. (\ref{eq:ChargeI}). There are two index current, proportional to $\partial_\mu\partial_i\varepsilon(x)$, and one index currents, proportional to $\partial_\mu \varepsilon(x)$ and    $ \partial_i\varepsilon(x)$. One index currents are not of interest here.  The two-index current is
\begin{eqnarray}
\propto i\,e^{i\,2\pi(\frac{I}{N}-\frac{I}{N})}\big(\psi ^*[Q_\psi \partial^\mu \partial^i \psi ]- [Q_\psi \partial^\mu \partial^i\psi ]^* \psi \big).
\end{eqnarray}
Let us consider the two-index phase at $I=1$ and $N=2$, \ie for $Q_\psi=1$ and $\psi=\frac{f+\rho}{\sqrt2 f}\,e^{i\,2\pi\phi/f}$,
\begin{eqnarray}
\frac{1}{2\pi i}j^{ij}=\frac{f}{M_{\rm eff}^2} \,\partial^i \partial^j\phi\to &&Q =\frac{f}{M_{\rm eff}^2}\int d^3 x\, \phi\,\partial^i \partial^j\phi  =\frac{f \lambda_{\rm thickness}}{M_{\rm eff}^2}\int d^2 x\,  \partial^i \partial^j\phi\\
&&\quad=   \,\sqrt{ \frac{\lambda v}{\mu}}\int\,  d\Omega~\frac{f\, v}{M_{\rm eff}^2}  j^{\theta\varphi}, \nonumber
\end{eqnarray}
where $v$ is the VEV of $\phi$ breaking $\Z_2$, $\mu$ is the $\phi$ mass as the result of this breaking, and $\lambda_{\rm thickness}$ is given in Eq. (\ref{eq:thickness}).  A delta function is obtained from the derivative of a step function in the angle direction. It is shown in   Fig. \ref{fig:TwoDdelta}. Here, the parameter $M_{\rm eff}$ is defined at the radius $r$ so that $Q$ becomese 1,
\begin{eqnarray}
M_{\rm eff}^2=   \sqrt{\lambda\,   \frac{f^2 v^3}{\mu}~},
\end{eqnarray}
and
\begin{eqnarray}
j^{\theta\varphi}(\theta,\varphi)=\frac{1}{r^2} \delta(\cos\theta-\cos\theta_0)
\delta(\varphi-\varphi_0).
\end{eqnarray}
With the current conservation, this surface integral is related to the head charge of the tadpole,
\begin{eqnarray}
\partial_\mu j^{\mu i}=0\to Q^i\propto \int_V d^3x\, \frac{\partial}{\partial x^j}\, j^{ji}=  \int_{\Sigma_{r}} d^2\sigma_{r}\,   j^{ri},\label{eq:Conservation}
\end{eqnarray}
where $\Sigma_{r}$ is the surface orthogonal to the radial direction. 
The current conservation used in (\ref{eq:Conservation}) is obtained via the equation of motion $\partial^2\psi=-m^2\psi$ and $\partial^2\psi^*=-m^2\psi^*$,
\begin{eqnarray}
\partial_\mu j^{\mu i}=\frac12\big[\psi^*\partial^i\partial^2 \psi +(\partial_\mu\psi^*)\partial^i\partial^\mu
\psi-(\partial^2\psi^*)\partial^i 
\psi- (\partial_\mu\psi^*)\partial^i\partial^\mu
\psi\big]=\frac12\big[\psi^*\partial^i\partial^2 \psi -(\partial^2\psi^*)\partial^i 
\psi \big]=0.
\end{eqnarray}
 
 \section{Conclusion}
Showing that discrete charges carry hairs in the bottom-up approach,  we  argued that discrete symmetries are  respected by gravity. It is based on the fact that discrete symmetries are  realized with domain walls  in the Universe  via the VEVs of Higgs fields, and the intersection of domain walls looks like a hair.  

\section*{\bf Appendix: Around a blackhole}

 \begin{figure}[!t]
\begin{center}
\includegraphics[width=0.55\linewidth]{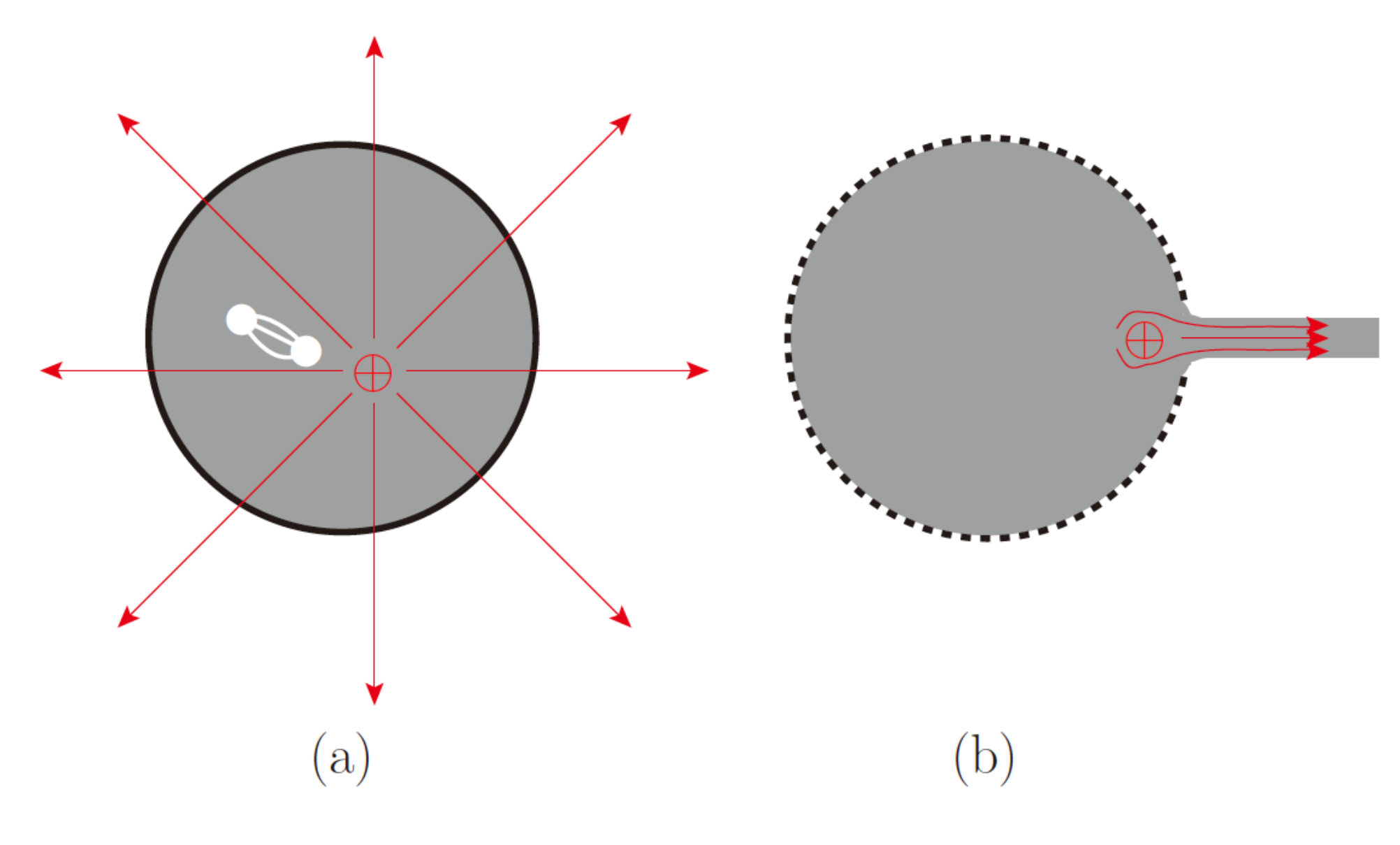} 
\end{center}
\caption{ A charged black hole. } \label{fig:ChBH}
\end{figure}

 \begin{figure}[!t]
\begin{center}
\includegraphics[width=0.55\linewidth]{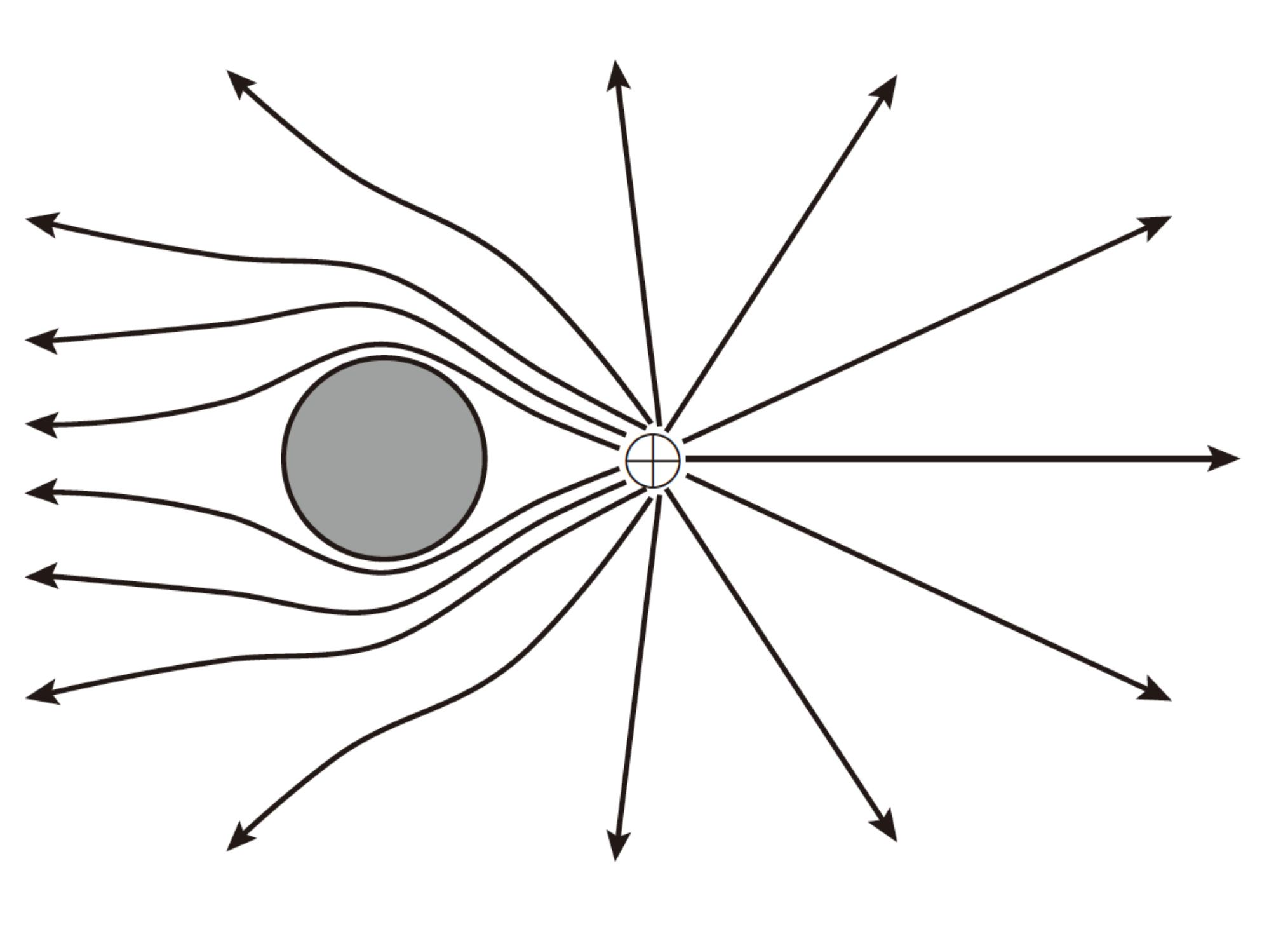} 
\end{center}
\caption{ A field vacua around a black hole. } \label{fig:BHvacOut}
\end{figure}
 
 Around a blackhole, the argument goes in parallel to the case of wormhole. So, we briefly point out the flux line argument in a charged Universe and draw the similarity in case of discrete charges.

The larger event horizon (out of two solutions) of the Reissner-Nordstr$\ddot{\rm o}$m black hole \cite{RNbh16} occurs at
\dis{
r_+=\frac12\left(r_S+ \sqrt{r_S^2-4r_Q^2} \right)\label{eq:RNhor}
}
where $r_S$ is the Schwarzschild radius without the charge. In Fig. \ref{fig:ChBH}\,(a), we show the situation. The event horizon takes into account the energy inside it, as depicted with the white illustration. It is basically the impossibility of the graviton field to go out of the horizon, bounded by $T_{\mu\nu}$. Any graviton field a spin-2 field,  can end at another energy point, due to the positive energy theorem,  inside the blackhole horizon. When we consider mass of a charged particle, it includes the field energy also. This count of energy subtracts the electromagnetic field energy permeating from the horizon to infinity, which is shown as $r_{Q}^2$ in Eq. (\ref{eq:RNhor}).  Within the closed boundary shown in Fig. \ref{fig:ChBH}\,(a), the electromagnetic field cannot end inside the blackhole horizon for a net charge $Q$ given inside the black hole. So, the field line goes out of the horizon of the gravity field, which looks like a hair from the black hole. Of course, the outside field is not included in $T_{\mu\nu}$ on the right-hand side of the Einstein equation. So, if fields permeating over the whole space are present, it is better to consider their effects just inside the blackhole. In Fig. \ref{fig:BHvacOut}  around a closed surface black hole, the electromagnetic field configuration, not interfering with the fact of the closed surface nature of the black hole, is illustrated.

\begin{figure}[!t]
\begin{center}
\includegraphics[width=0.55\linewidth]{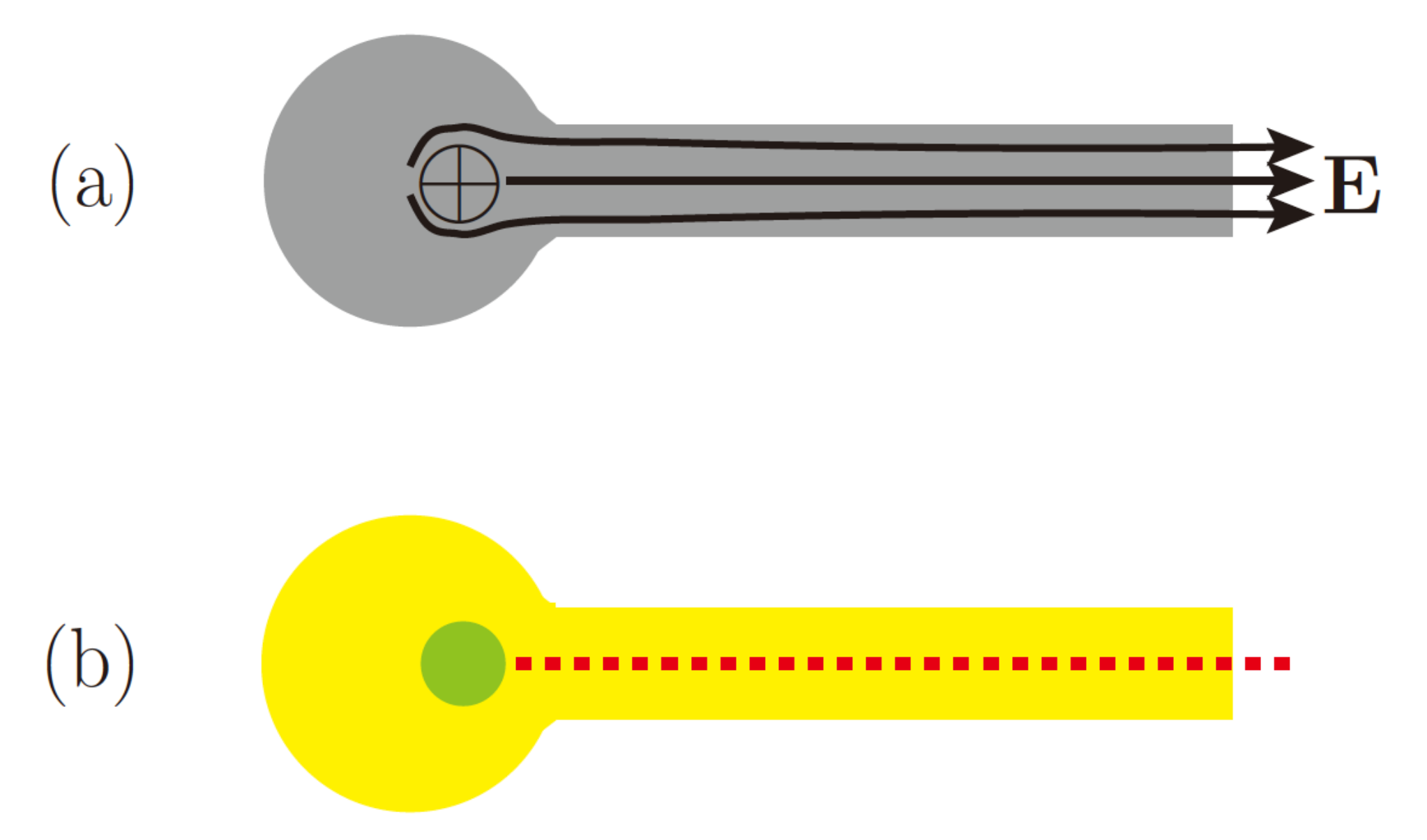} 
\end{center}
\caption{ Hairs of a black hole: (a) for an electromagnetic charge, and (b) for a discrete tadpole. } \label{fig:HairsBH}
\end{figure}

Now, let us consider that the boundary of the fields is confined inside the horizon.   If the field lines are forbidden to cross the boundary for the spin-1 fields also, one cannot allow a net charge $Q$ inside the horizon and must give up the closed surface of the blackhole,\footnote{In the closed Universe, a non-zero charge must break the U(1) gauge symmetry \cite{TLee93}.} and the metric must allow the open geometry as shown in Fig. \ref{fig:ChBH}\,(b). By gauge transformation, the electromagnetic fields of \ref{fig:ChBH}\,(a) are transformed to those of Fig. \ref{fig:ChBH}\,(b) where, in the most part of the blackhole space, the gravity field is bounded within the dashed area but the electromagnetic field goes out into the open space through the pinched hole. This is the hair we explain within our set-up. Spin-0 fields do not have the flux lines but can have different quantum numbers distinguishing different domains of the spin-0 field vacua. For spin-1 field the hair is the field strength and for spin-0 field we argued that it must be the intersection of domain walls. This was named as `tadpole'. In Fig. \ref{fig:HairsBH}, the hairs of U(1) gauge field and $\Z_2$ tadpole are shown.  In Fig. \ref{fig:HairsBH}\,(b), the discrete vacuum of the blackhole in the open geometry is $q=0$.
Here, we note the difference between the flux line  and the discrete tadpole. The electromagnetic flux carries energy and hence affects the blackhole radius. But, the discrete vacua in Fig. \ref{fig:Vacua} are degenerate and different vacua have the same energy. If they have strings (intersection of domain walls), they must be taken into account in the energy calculation inside the blackhole, not considering the outside part. So, the blackhole radius is not changed.

 \begin{figure}[!t]
\begin{center}
\includegraphics[width=0.4\linewidth]{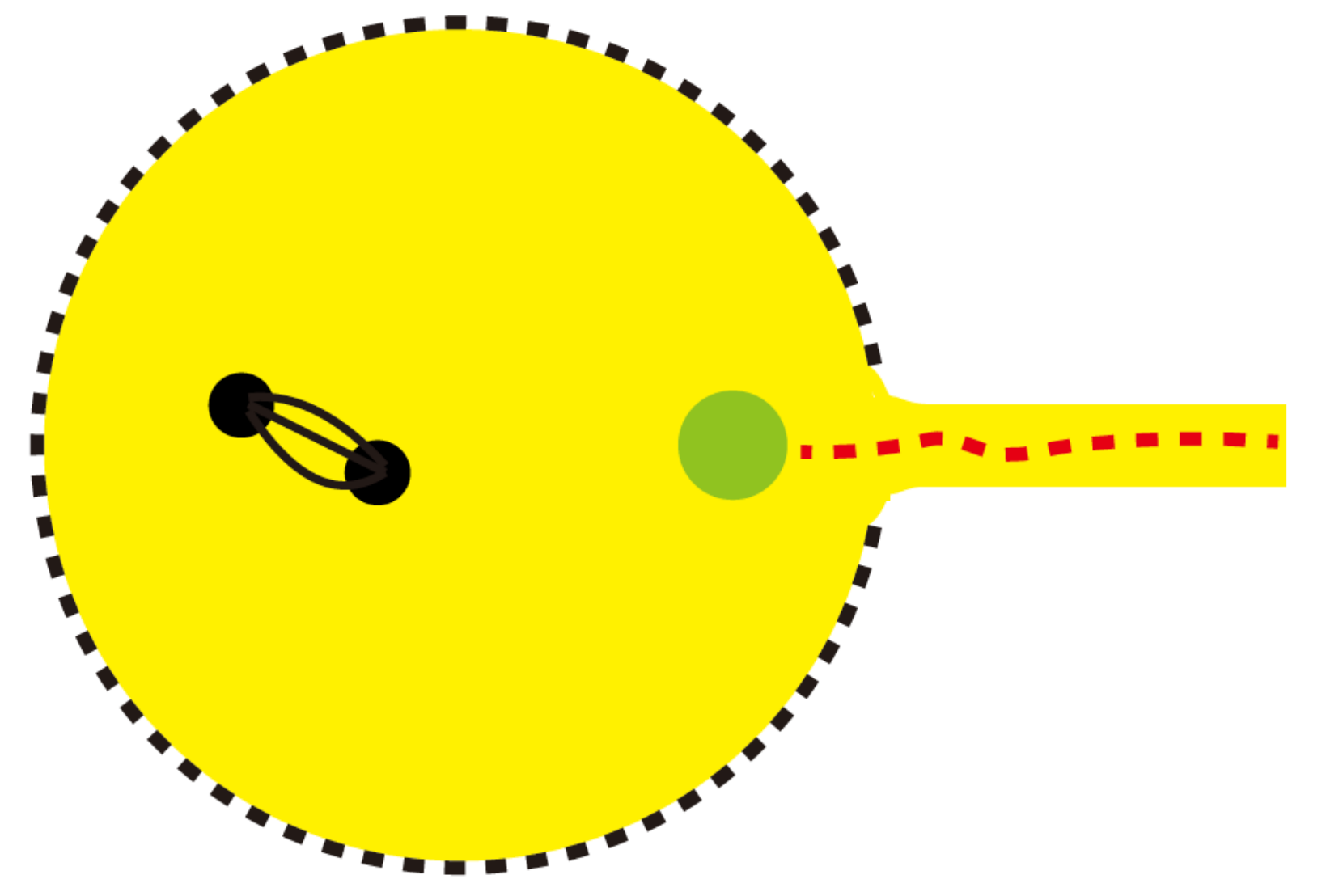} 
\end{center}
\caption{A tadpole is thrown into the $q=0$ vacuum. } \label{fig:BHvacIn}
\end{figure}

Vacua of spin-0 field are not constrained by the blackhole horizon, and the same vacuum can be connected to the outside as the field lines of  Fig. \ref{fig:ChBH}\,(a). In Fig. \ref{fig:BHvacIn}, a $\Z_2$ tadpole is thrown into the $q=0$ blackhole vacuum, and the outside observer notices that he lost a discrete charge $q=1$ to the blackhole.  Namely, he notices that the black hole has a $\Z_2$ hair. If the tail is cut, the outside observer notices that the blackhole ate only even number of $\Z_2$ charge with some energy increase inside the blackhole but the discrete charge is not increased in the blackhole. Thus, the existence of a blackhole does not violate the discrete symmetry.
 
\acknowledgments{ 
J.E.K. is supported in part by the National Research Foundation (NRF) grant funded by the Korean Government (MEST) (NRF-2015R1D1A1A01058449) and  the IBS (IBS-R017-D1-2016-a00), and B.K.  is supported in part by the NRF-2013R1A1A2006904.  }

\end{document}